\documentstyle[pre,aps,multicol,epsf]{revtex}

\draft

\begin{document}

\title{Branching and annihilating L\'evy flights}

\author{Daniel Vernon and Martin Howard}

\address{Department of Physics, Simon Fraser University, Burnaby,
         British Columbia, Canada V5A 1S6.}

\date{November 28, 2000}

\maketitle

\begin{abstract}
We consider a system of particles undergoing the branching and
annihilating reactions $A\to (m+1)A$ and $A+A\to\emptyset$, with $m$
even. The particles move via long--range L\'evy flights, where the
probability of moving a distance $r$ decays as $r^{-d-\sigma}$. We
analyze this system of {\em branching and annihilating L\'evy flights}
(BALF) using field theoretic renormalization group techniques close to 
the upper critical dimension $d_c=\sigma$, with $\sigma<2$. These
results are then compared with Monte--Carlo simulations in $d=1$. For
$\sigma$ close to unity in $d=1$, the critical point for the
transition from an absorbing to an
active phase occurs at zero branching. However, for $\sigma$ bigger than
about $3/2$ in $d=1$, the critical branching rate moves smoothly away
from zero with increasing $\sigma$, and the transition lies in a
different universality class, inaccessible to controlled perturbative
expansions. We measure the exponents in both universality classes and
examine their behavior as a function of $\sigma$.
\end{abstract}



\begin{multicols}{2}
\narrowtext


\section{Introduction}

Systems possessing a continuous nonequilibrium phase transition from
an active into an empty, absorbing state have been intensively studied
in the past few years. Despite the wide variety of processes that have
been investigated, it has proved possible to classify the critical
properties of these transitions into a small number of universality
classes. Although the well known case of directed percolation (DP) 
\cite{DP,moredp,hayerev} has turned out to be the most common
universality 
class, many investigations have examined systems with quite different
critical properties. For instance, the model of {\it branching
and annihilating random walks} with an {\em even} number of offspring
(BARW) defines a separate universality class
\cite{hayerev,tt,jensen:1994,cardy-tauber}. This reaction--diffusion
system 
consists of random walkers able to undergo the branching and
annihilating reactions $A\to (m+1)A$ and $A+A\to\emptyset$, with $m$
even. Other models in this class (at least in $d=1$) 
include certain probabilistic cellular automata \cite{automata},
monomer--dimer models \cite{monomerdimer,hwang-etal:1998,inf_DP_DI},
nonequilibrium kinetic Ising models \cite{menyhard-odor}, and
generalized DP with two 
absorbing states \cite{haye}. These models escape from the DP
universality class by possessing an extra conservation law or
symmetry. The BARW model respects an additional ``parity''
conservation of the total number of particles modulo $2$. On the other
hand, branching and annihilating random walks with an odd number of
offspring possess no such ``parity'' conservation, and hence
belong to the DP universality class\cite{cardy-tauber}. For the
other models mentioned above 
\cite{automata,monomerdimer,hwang-etal:1998,inf_DP_DI,menyhard-odor,haye},
the DP class is escaped via an underlying symmetry between the
absorbing states. 

Both the DP and BARW classes do, however, share one important feature:
the dynamical processes involved are short--ranged. One would expect
that the addition of long--ranged processes would significantly alter the properties of the
active/absorbing transitions. Recently this expectation was confirmed
by investigations of L\'evy DP (LDP). This modification, originally
proposed by Mollison \cite{mollis} in the context of epidemic
spreading, is a generalization of DP where the distribution of
spreading distances $r$ is given by
\begin{equation}
P(r)\sim 1/r^{d+\sigma},\qquad(\sigma>0),
\label{levydef}
\end{equation}
where $d$ is the spatial dimension of the system, and $\sigma$ is a
free parameter (the L\'evy index) that controls the characteristic
shape of the distribution. This distribution is asymptotically (as
$r\to\infty$) equal to a L\'evy distribution, and we will loosely refer to
it as such. It was first suggested that the
critical exponents describing the LDP transition should vary
continuously with $\sigma$ \cite{grassb}. This expectation was backed
up by field theoretic renormalization group calculations in
Ref.~\cite{janssn}, and confirmed numerically in
Refs.~\cite{mcm,hinhow}. Note that other numerical
work~\cite{albano,cannas} introduced an upper cut off for the
flight distance $r$. This resulted in effective
{\em short--range} behavior, meaning that the LDP regime was not properly
accessed. The results of Ref.~\cite{cannas} also appear to be
adversely affected by strong finite size effects.

The purpose of the present paper is to further investigate the impact
of L\'evy flights in models with nonequilibrium phase transitions.
We will analyze in detail a model of {\em branching and annihilating
L\'evy flights} with an {\em even} number of offspring (BALF), a
straightforward generalization of the BARW
model, where the random walkers are replaced by particles performing
L\'evy flights. The BALF model possesses an upper
critical dimension $d_c$ which varies continuously with the L\'evy
index $\sigma$. For $d<d_c$, the model contains two new
universality classes resulting from the long--range nature of the
L\'evy flights. The exponents in both of these classes also vary
continuously with $\sigma$. We will investigate these new universality
classes using field theoretic methods, some exact results
for the pure annihilation model (where the branching parameter is set
equal to zero), and Monte--Carlo simulations in $d=1$.

A further attractive feature of the BALF model is that it casts some
additional light on the properties of the ordinary short--ranged BARW
model. We will
see that changing the L\'evy index from $\sigma=1$ to $\sigma=2$ for
the BALF model in fixed dimension $d=1$ is in
many ways similar to changing the physical dimension from $d=2$ to
$d=1$ in the short--ranged BARW model. Although this correspondence is
certainly not rigorous, we can nevertheless use simulations of the
BALF model in the physical dimension $d=1$ to better understand
properties of the BARW model which lie in the inaccessible
dimensions between $d=1$ and $d=2$. This will allow us to probe
numerically some important features of the BARW field theory developed
by Cardy and T\"auber in Ref.\cite{cardy-tauber}.

We now give a brief summary of the layout of the paper. In the next
section we briefly review the relevant properties of the short--ranged
BARW model. In Section~III we then introduce the BALF model and
present its mean field behavior. After these preliminaries we then
present the field theoretic action for BALF, which we analyze using
diagrammatic and renormalization group methods. These results are then
compared with Monte--Carlo simulations in Section~IV. Finally, our
conclusions appear in Section~V.

\section{Branching and annihilating random walks}

The BARW model is defined by the following reaction processes:
\begin{eqnarray}
& A\to (m+1)A, \qquad & {\rm rate}~\mu_m, \\
& A+A\to\emptyset, & {\rm rate}~\lambda,
\end{eqnarray}
where the identical particles $A$ otherwise perform simple random
walks with diffusion constant $D$. As the reaction rate parameters are
varied, one finds a continuous phase transition from a region
controlled by the pure annihilation process to an active region
characterized by a nonzero particle density in the steady--state. 
The growth of BARW clusters close to the critical point can be
summarized by a set of independent exponents. A natural choice is to
consider $\nu_\perp$ and $\nu_\parallel$, which describe the divergence
of the  correlation lengths in space, 
$\xi_\perp \sim |\Delta|^{-\nu_\perp}$, and time, $\xi_\parallel
\sim |\Delta|^{-\nu_\parallel}$, close to criticality. Here the
parameter $\Delta$ describes the deviation from the critical point at
the active/absorbing transition. 
We also need the order parameter exponent $\beta$, which can 
be defined in two {\it a priori} different ways: it is either governed by 
the probability that a cluster grown from a finite seed never dies,
\begin{equation}
\label{P_bulk}
         P(t\to\infty,\Delta) \sim \Delta^{\beta_{\rm seed}}, \qquad
         \Delta > 0, 
\end{equation}
or by the coarse-grained density of active sites in the steady state,
\begin{equation}
\label{n(Delta)}
         n(\Delta) \sim \Delta^{\beta_{\rm dens}}, \qquad \Delta > 0.
\end{equation}
These exponents can be simply calculated in mean field theory, valid
for $d>d_c=2$. The appropriate mean field rate
equation for the coarse--grained density $n({\bf x},t)$ is given by
\begin{equation}
\partial_t n=D\nabla^2 n + m \mu_m n - 2 \lambda n^2 .
\label{mfeq}
\end{equation}
For $\mu_m=0$ no branching is present, and we are reduced to the
well known annihilation reaction $A+A\to\emptyset$, which
asymptotically exhibits a power law 
mean field density decay $n\sim t^{-1}$. However, for
nonzero $\mu_m$, we have the homogeneous steady--state solution
$n_s=m\mu_m/2\lambda$. Hence the critical value of $\mu_m$
clearly lies at zero, and we identify $\Delta=m\mu_m$.
The density thus behaves as $n_s\propto\Delta$, and we immediately
see that $\beta^{MF}_{\rm dens}=1$. The alternative order parameter 
exponent 
$\beta_{\rm seed}^{MF}$ can also be simply calculated: for $d>d_c=2$,
the survival probability (\ref{P_bulk}) of a particle cluster will be
finite for {\it any\/} value of the branching rate, implying that
$\beta^{MF}_{\rm seed}=0$. This result follows from the
non--recurrence of random walks in $d>2$. The correlation length
exponents can also be simply derived from Eq.~(\ref{mfeq}), yielding
$\nu_{\perp}^{MF}=1/2$, $\nu_{\parallel}^{MF}=1$. Hence the dynamic
exponent $z$, defined by $\xi_{\perp}\sim\xi_{\parallel}^{1/z}$,
is given by $z^{MF}=\nu_{\parallel}^{MF}/\nu_{\perp}^{MF}=2$. 

Below the upper critical dimension the above mean field
analysis breaks down due to the presence of fluctuations. Recently,
methods have been developed to systematically include these 
fluctuation effects. Firstly, the appropriate master equation, which
provides a complete description of the microscopic dynamics of the
system, is transformed into a second--quantized Hamiltonian. This
representation is then mapped onto a coarse--grained field theoretic
action \cite{peliti,lee,cardy}. From this point the standard tools of
renormalized perturbation expansions can be
employed, and the effects of fluctuations systematically computed. For
the case of BARW, this analysis was performed in
Ref.~\cite{cardy-tauber}. In the following we summarize the main
results of that analysis. The field theoretic action for BARW, written
in terms of the response field $\hat\psi({\bf x},t)$ and the  
``density'' field $\psi({\bf x},t)$, is given by \cite{cardy-tauber}
\begin{eqnarray}
& & S_0[\psi,\hat\psi;\tau]=
\int d^dx \bigg[ \int_0^{\tau} dt \Big[\hat\psi({\bf
x},t)[\partial_t- D{\bf\nabla}^2]\psi({\bf x},t) \nonumber \\
& & \qquad\qquad\qquad\qquad\qquad\quad
-\lambda[1-\hat\psi({\bf x},t)^2]\psi({\bf x},t)^2
\label{baweaction} \\
& & \qquad\qquad\qquad\qquad\quad + \mu_m[1-\hat\psi({\bf
x},t)^m]\hat\psi({\bf x},t)\psi({\bf x},t)\Big]  \nonumber \\
& & \qquad\qquad\qquad\qquad\qquad\quad
-\psi({\bf x},\tau)-n_0\hat\psi({\bf x},0)\bigg] \nonumber .
\end{eqnarray}
Here the terms on the first line of (\ref{baweaction}) represent
diffusion of the particles (with continuum diffusion constant
$D$). The second line describes the annihilation reaction (with
continuum rate $\lambda$), while the terms on the third line
represent the branching process (with continuum rate $\mu_m$). The
final two terms represent, respectively,  a contribution due to the
projection state (see Ref.~\cite{peliti}), and the initial condition
(an uncorrelated Poisson distribution with mean $n_0$). In the
following we will restrict ourselves to the case of {\it even} $m$,
since it is known that the odd $m$ case belongs to the DP universality
class \cite{cardy-tauber}.

The action given in (\ref{baweaction}) is a {\it bare}
action. In order to properly include fluctuation effects one must be 
careful to include processes generated by a combination of branching
and annihilation. In other words in addition to the process $A\to
(m+1)A$, the reactions $A\to (m-1)A$, \ldots, $A\to 3A$ need to be
included. These considerations lead to the full action
\begin{eqnarray}
& & S[\psi,\hat\psi;\tau]=
\int d^dx  \bigg[\int_0^{\tau} dt \Big[\hat\psi({\bf x},t)
[\partial_t- D{\bf\nabla}^2]\psi({\bf x},t) \nonumber \\
& &\qquad\qquad\qquad + \sum_{l=1}^{m/2}\mu_{2l}[1-\hat\psi({\bf x},t)^{2l}]
\hat\psi({\bf x},t)\psi({\bf x},t) \label{barwbulk} \\
& &  \qquad\quad -\lambda[1-\hat\psi({\bf x},t)^2]\psi({\bf x},t)^2\Big]
-\psi({\bf x},\tau)-n_0\hat\psi({\bf x},0)\bigg] \nonumber .
\end{eqnarray}
Notice also that (for {\it even $m$}) the action (\ref{barwbulk})
is invariant under the ``parity'' transformation
\begin{equation}
\label{parity}
\hat\psi({\bf x},t)\to -\hat\psi({\bf x},t), \qquad \psi({\bf x},t)\to
-\psi({\bf x},t) .
\end{equation}
This symmetry corresponds physically to particle conservation modulo
$2$. The presence of this extra symmetry now takes the system away
from the DP universality class, and into a new class: that of
branching and annihilating random walks with an even number of offspring.

Simple power counting on the action in Eq.~(\ref{barwbulk}) reveals
that the upper critical dimension is $d_c=2$. Close to $d_c$,
the renormalization of the above action is
quite straightforward (here we again quote the results from
Ref.~\cite{cardy-tauber}). At the annihilation fixed point the RG
eigenvalue of the branching parameter can easily be computed. To one
loop order one finds
$y_{\mu_m}=2-m(m+1)\epsilon/2+O(\epsilon^2)$, where $\epsilon=2-d$.
Hence we see that the {\it lowest} branching process is
actually the most relevant. Therefore, close to $2$ dimensions where
the branching remains relevant, we expect to find an {\it active}
state for all nonzero values of the branching (in agreement with
the mean field theory presented above). 
Furthermore, in this regime, we can exploit the fact that the
critical point, which remains at zero branching, is described by the
pure annihilation theory. Matching the exactly known density decay~\cite{lee}
and survival probability exponents in the annihilation theory with
their counterparts in the critical BARW theory yields the exact
exponent relations $\beta_{\rm dens}=d\nu_{\parallel}/2$,
$\beta_{\rm seed}=\nu_{\parallel}(2-d)/2$, and $z=2$
\cite{cardy-tauber}. To the best of
our knowledge, the result for $\beta_{\rm seed}$, although simple to
derive, has not previously been given in the literature. 

Inspection of the one loop result for the most relevant RG
eigenvalue $y_{\mu_2}$ shows that it eventually becomes negative. This
occurs at a second critical dimension $d_c'$, where $d_c'=4/3$ to one
loop order. For $d<d_c'$ we expect a 
major change in the behavior of the system, since the branching
process will no longer be relevant at the annihilation fixed
point. The critical transition point is then
shifted with the active state only being present for values of the
branching greater than some positive critical value. For branching
parameter values smaller than this value, the branching will be
asymptotically irrelevant. This region of parameter space will thus be
controlled by the annihilation fixed point of the $A+A\to\emptyset$
process, where the density decays away as a power law. Hence this
region of parameter space should be considered a 
critical inactive (or absorbing) phase. The presence of a second
critical dimension $d_c'\approx 4/3$ immediately rules out any
possibility of using perturbative $\epsilon$ expansions to access 
the non--trivial active/absorbing transition expected in the
physical dimension $d=1$. Instead cruder 
techniques (such as the loop expansion in fixed dimension) must be
employed \cite{cardy-tauber}. We will not discuss this part of the
analysis of Ref.~\cite{cardy-tauber} in much
detail. However we do wish to point out that the truncated loop
expansion at one loop does predict a jump in the critical point at
around $d_c'\approx 4/3$, from zero branching to some finite value. We
will have more to say about his observation in Section~IV, after we
have presented our analytical and numerical study of the BALF model. 

It is also possible to analyze the BARW model in $d=1$ using exact
methods. In Ref.~\cite{cardy-tauber} it was demonstrated that, at the
annihilation fixed point, the one loop RG eigenvalue $y_{\mu_2}$ is
actually exact in $d=1$. However, the reason for the cancellation of
the contributions from higher loop orders in the field theory remains
unclear. Other work \cite{schutz,hfl}, using quantum
spin Hamiltonians, has indicated that the exponents $\beta_{\rm dens}$
and $\beta_{\rm seed}$ are exactly equal at the active/absorbing
transition in $d=1$. This conclusion is also supported by numerical
simulations \cite{jensen:1997}.

\section{Branching and annihilating Levy flights}

We now turn to the main object of this paper: a systematic
investigation of the BALF model. To begin with, we consider the model
at the mean field level. The appropriate mean field equation is
given by 
\begin{equation}
\partial_t n=(D_N\nabla^2 + D_A\nabla^{\sigma}) n + m \mu_m n - 2
\lambda n^2 ,
\label{mflevyeq}
\end{equation}
where $D_N$ and $D_A$ are the rates for normal and anomalous (L\'evy)
diffusion, respectively. The anomalous diffusion operator
$\nabla^{\sigma}$ describes moves over long distances and is defined
by its action in momentum space 
\begin{equation}
\nabla^{\sigma}e^{i{\bf k}\cdot{\bf x}}=-k^{\sigma}e^{i{\bf
k}\cdot{\bf x}} ,
\label{levyk}
\end{equation}
where $k=|{\bf k}|$. The standard diffusion term $D_N\nabla^2$ takes
into account the short--range component of the L\'evy distribution.
A more detailed derivation and justification for the L\'evy term can
be found in Ref.~\cite{hinhow}. The mean field exponents can now easily
be extracted. The critical point remains at zero branching, 
and, for $\sigma<2$, we identify $\beta_{\rm dens}^{MF}=1$,
$\beta_{\rm seed}^{MF}=0$, 
$\nu_{\parallel}^{MF}=1$, and $\nu_{\perp}^{MF}=1/\sigma$. Note that,
for $\sigma>2$, these exponents cross over smoothly to the
ordinary mean field BARW exponents. Even at the mean field level, we
can see that the exponent $\nu_{\perp}$ varies continuously with the
L\'evy index $\sigma$. 

The above mean field description will only be quantitatively valid
above the upper critical dimension. For $d\leq d_c$, we must again
take fluctuation effects into account. This can be done using the same 
methods as were used for the short--ranged BARW model
\cite{cardy-tauber}. We emphasize that the inclusion of the 
long--ranged L\'evy processes does not introduce any particular
difficulties 
for the field theory mapping (see Ref.~\cite{hinhow} for further
details). Specializing immediately to the case with $m=2$, and
defining $\mu \equiv \mu_2$, we find that the field theoretic action is given
by 
\begin{eqnarray}
& & \qquad\qquad\qquad\qquad\qquad S[\psi,\hat\psi;\tau]= \nonumber
\\ & & \quad
\int d^dx \bigg[ \int_0^{\tau} dt \Big[\hat\psi
({\bf x},t)[\partial_t- 
D_N{\bf\nabla}^2 -D_A{\bf\nabla}^{\sigma}]\psi({\bf x},t)
\nonumber \\ 
& &\qquad\qquad\qquad\qquad\quad
-\lambda[1-\hat\psi({\bf x},t)^2]\psi({\bf x},t)^2
\label{balfeaction} \\
& &\qquad\qquad\qquad\quad + \mu[1-\hat\psi({\bf x},t)^2]\hat\psi({\bf 
x},t)\psi({\bf x},t)\Big] \nonumber \\
& & \qquad\qquad\qquad\qquad\quad
 -\psi({\bf x},\tau)-n_0\hat\psi({\bf x},0)\bigg] \nonumber .
\end{eqnarray}
This action describes both normal and anomalous diffusion.
The naive scaling dimensions of the fields are
\begin{equation}
        [ {\hat \psi}({\bf x},t) ] = \kappa^0 \, , \quad
        [ \psi({\bf x},t) ] = \kappa^d \ .
\end{equation}
With $[x]=\kappa^{-1}$ and $[t]=\kappa^{-\sigma}$, we see that the
naive scaling dimensions of the couplings are
\begin{equation}
        [D_N] = \kappa^{\sigma-2} \, , \quad [D_A] = \kappa^0 \, , \quad
        [\lambda] = \kappa^{\sigma-d} \, , \quad [ \mu ] = \kappa^{\sigma} \ .
\end{equation}
Hence, power counting reveals that the upper critical dimension, at
which the fluctuations become important, is $d_c=\sigma$, for
$\sigma<2$. 

\begin{figure}
\centerline{\epsfxsize=4cm 
\epsfbox{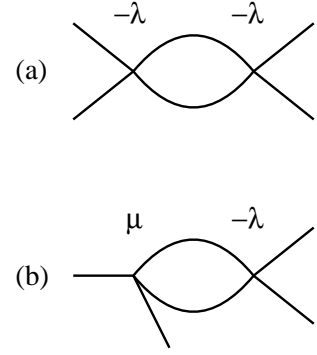}
}
\caption{One loop contribution to (a) the renormalized annihilation
vertex and (b) the renormalized branching vertex.}
\label{vertices}
\end{figure}

We have calculated the renormalization group flow functions and
eigenvalues, so as to determine the long distance and late time
behavior of this field theory.  The one loop contribution to the
renormalized annihilation vertex is given by the diagram in
Fig.~\ref{vertices}a. For the case where $\mu=0$, the propagator is
$(s +D_N k^2 + D_A k^{\sigma})^{-1}$ in $({\bf k},s)$ space ($s$
is the Laplace transformed time variable), or
$e^{-(D_Ak^{\sigma}+D_Nk^2)t}$ in $({\bf k},t)$ space. It turns out to
be easiest to calculate an extended--time vertex function in $({\bf
k},t)$ space, and then determine the renormalized coupling in $({\bf
k},s)$ space by performing a Laplace transform and evaluating at
the normalization point $({\bf k},s)=({\bf 0}, 2D_A\kappa^{\sigma})$. 

The first step is to drop the normal diffusion term, as it is less
relevant for $\sigma < 2$.  
The dimensionless renormalized annihilation coupling is defined by
\begin{equation}
\ell = Z_{\lambda} \lambda C_d \kappa^{-\epsilon}/D_A , 
\end{equation}
with $\epsilon=d_c-d=\sigma-d$, and 
\begin{equation}
C_d ={{\Gamma(d/\sigma)}\over{\Gamma(d/2)}}
{{\Gamma(2-d/\sigma)}\over{2^{d-1}\pi^{d/2}}}. 
\end{equation}
The one loop renormalization factor $Z_{\lambda}$ is then
\begin{equation}
Z_{\lambda} = 1 - {\lambda\over D_A} \frac{C_d\kappa^{-\epsilon}}
{\epsilon}.
\end{equation}
Hence the $\beta$ function is given by
\begin{equation}
\beta_{\ell}(\ell) \equiv \kappa {{\partial \ell} \over {\partial
\kappa}}= \ell(d-\sigma+\ell) ,
\label{betaell}
\end{equation}
with fixed points at $\ell=0$ and $\ell =\ell^{*}
=\epsilon=\sigma-d$. The result (\ref{betaell}) is actually exact to
all orders in perturbation theory \cite{lee}. For $d>\sigma$, the
Gaussian fixed point at $\ell=0$ is stable, while for $d < \sigma$,
the non--trivial $O(\epsilon)$ fixed point at $\ell=\ell^{*}$ is
stable. 

To investigate the relevance of the branching process, we now
calculate the one loop RG eigenvalue for the branching process at the
annihilation fixed point. Defining the dimensionless renormalized
branching rate as
\begin{equation}
s = Z_{\mu} \mu \kappa^{-\sigma}/D_A ,
\end{equation} 
then, from the diagram in Fig.~\ref{vertices}b, we can compute the one
loop renormalization factor
\begin{equation}
Z_{\mu} = 1 - 3 {{\lambda} \over {D_A}} {{C_d \kappa^{-\epsilon}} \over
{\epsilon}} .
\end{equation}
Hence the $\zeta$ function is
\begin{equation}
\zeta_{\mu} \equiv \kappa {{\partial} \over {\partial \kappa}} \ln
{{s} \over {\mu}} = -\sigma + 3 \ell + {\mathcal O}(\ell^2).
\end{equation}
Thus the one loop RG eigenvalue for the branching process at the
annihilation fixed point is 
\begin{equation}
\label{ymu}
y_{\mu} = -\zeta_{\mu}(\ell^*) = \sigma - 3 \epsilon = 3d -2\sigma.
\end{equation}
Consequently, according to the one loop theory, the
branching process is relevant at the annihilation fixed point for 
\begin{equation}
\sigma < \sigma_c'(d)=3d/2,
\label{sigmac}
\end{equation}
or, in $d=1$, for $\sigma < \sigma_c'(d=1)=3/2$. Hence, as in the mean
field case, we expect
an active phase for all nonzero values of the branching rate $\mu$,
for sufficiently small $\sigma$ (see also the phase diagram in section
IV). In this regime, we can again exploit the fact that criticality
lies at zero branching, and hence that the
critical behavior of the BALF model coincides with that for the simple
L\'evy annihilation model $A+A\to\emptyset$ \cite{hinhow}. For the
L\'evy annihilation
model, several exact results can be derived: the density decays as
$t^{-d/\sigma}$ (for $d<\sigma<2$) \cite{hinhow}; the survival
probability decays as $t^{d/\sigma-1}$ (also for $d<\sigma<2$); and
the dynamic exponent is just $z=\sigma$ (for $\sigma<2$). The
second of these results follows in a simple way from the analysis of
Ref.~\cite{hughes}, but is nevertheless, to the best of our knowledge,
a new result. On the other hand in the BALF model it is
straightforward to show that, at criticality, the density should decay
as $t^{-\beta_{\rm dens}/\nu_{\parallel}}$, and the survival
probability as $t^{-\beta_{\rm seed}/\nu_{\parallel}}$, and where
again we have $z=\sigma$ at the L\'evy annihilation fixed point
\cite{hinhow}. Matching these results to the L\'evy annihilation case,
we have $\beta_{\rm dens}=d\nu_{\parallel}/\sigma$ and 
$\beta_{\rm seed}=\nu_{\parallel}(\sigma-d)/\sigma$.

Hence, in the regime $\sigma<\sigma_c'(d)$, there is just one
independent exponent which
must be calculated perturbatively. Following a similar
analysis to that in Ref.~\cite{cardy-tauber}, this exponent can be
taken to be $\nu_{\perp}$, which in terms of the RG eigenvalue for the
branching is given by $\nu_{\perp}=1/y_{\mu}$. Ideally, at the
annihilation fixed point in $d=1$, one would like to be able to
calculate the RG eigenvalue $y_{\mu}$ exactly, as was done in the
short--ranged BARW model \cite{cardy-tauber}. Remarkably in that case
it was found 
that the one loop result was exact. Unfortunately, generalizing the
methods of Ref.~\cite{cardy-tauber} to the L\'evy case does not seem
to be straightforward. However there is some numerical evidence (to be
presented in the next section) to suggest that the one loop result for
$y_{\mu}$ in the L\'evy case is again exact.

We also note that, in the above regime, if $\sigma=2<\sigma_c'$ 
then, to one loop order, we expect a smooth crossover to the
short--ranged BARW model. Hence, at
least at the one loop level, the model in this regime is 
more straightforward than the LDP case, where there are additional
complications (see Refs.~\cite{janssn,hinhow} for more details).

We now discuss the case where, for $\sigma > \sigma_c'(d)$,
the branching becomes irrelevant at the L\'evy annihilation fixed
point. For this regime to be present at all, then from
Eq.~(\ref{sigmac}), we require $d<4/3$ to one loop order. In this
regime we expect the critical branching rate $\mu_c(\sigma,d)$ to
become nonzero. For $0<\mu<\mu_c(\sigma,d)$, the branching will be
asymptotically irrelevant, and this phase will again be governed by
the exponents of the pure L\'evy annihilation universality class
\cite{hinhow}. At $\mu=\mu_c(\sigma,d)$ we then expect a non--trivial
transition to an active phase. As was the case for the short--ranged
BARW model we expect this transition to be inaccessible to
controlled perturbative expansions, and in a different universality
class to that discussed above. This follows from the fact that
this transition only appears below a second critical dimension
$d<d_c'(\sigma)= 2\sigma/3$ to one loop order. Hence, as was the case
for the short--ranged BARW model, $\epsilon=d_c-d$ expansions down
from the upper critical dimension $d_c=\sigma$ will not be able to
access this transition.  

We note that a ``precursor'' of the critical inactive phase present
for $\sigma>\sigma_c'(d)$ is already
evident in the $\sigma<\sigma_c'(d)$ regime. Using the above
analysis we see that $\beta_{\rm dens}=d/y_{\mu}=d/(3d-2\sigma)$
for $\sigma<\sigma_c'(d)$ (to one loop). Hence, to this order, as
$\sigma\to\sigma_c'(d)=3d/2$ from below, $\beta_{\rm dens}$ 
diverges. This implies that for a fixed small value of the
branching, the active phase has a decreasing density as a function of
$\sigma$, as $\sigma$ is increased towards $\sigma_c'(d)$. Finally at
$\sigma=3d/2$ (to one loop order) an entire critical inactive phase
opens up.  

We can now see the similarities between properties of the
short--ranged 
BARW model as dimension is lowered from $d=2$ to $d=1$, and the BALF
model in $d=1$, as the L\'evy index is raised from $\sigma=1$ to
$\sigma=2$. In particular, to one loop order, the region
$1<\sigma<\sigma_c'(d=1)=3/2$ 
for the $d=1$ BALF model contains the direct analog of the
inaccessible universality class present in BARW for $d_c'\approx
4/3<d<2$. 

\section{Simulation results}

In order to further investigate the BALF universality class, we have
performed extensive numerical simulations of a lattice BALF model in
$d=1$. At each time step, a randomly chosen particle was allowed
either to branch, with probability $1-p$, or to move, via a
long--range jump, with probability $p$;  $p$ was the only parameter in
the simulations. 
The number of particles at each lattice site was restricted to zero or 
one: thus, when a particle moved to an occupied site, both particles
were annihilated. At each branching step, a particle produced two
offspring, which occupied the two sites
to the immediate left or right of the original particle, with the
side chosen randomly. As pointed out in Ref.~\cite{kwon-park:1995},
this method of choosing occupied sites is necessary since if the newly
occupied sites are chosen symmetrically about the original site, then
the short--ranged BARW model turns out to be in its inactive 
state for all $0<p\leq 1$.

The distribution of hop lengths was chosen to follow
Eq.~(\ref{levydef}), for $r\ge 1$.  This distribution is 
implemented by choosing a random number $x$ from a uniform
distribution on the interval $[0,1)$, and then calculating a new
random variable $r=(1-x)^{-1/\sigma}$. It is easy to see that this 
produces a sequence of numbers whose distribution follows
Eq.~(\ref{levydef}). 

\begin{figure}
\centerline{\epsfxsize=8.5cm
\epsfbox{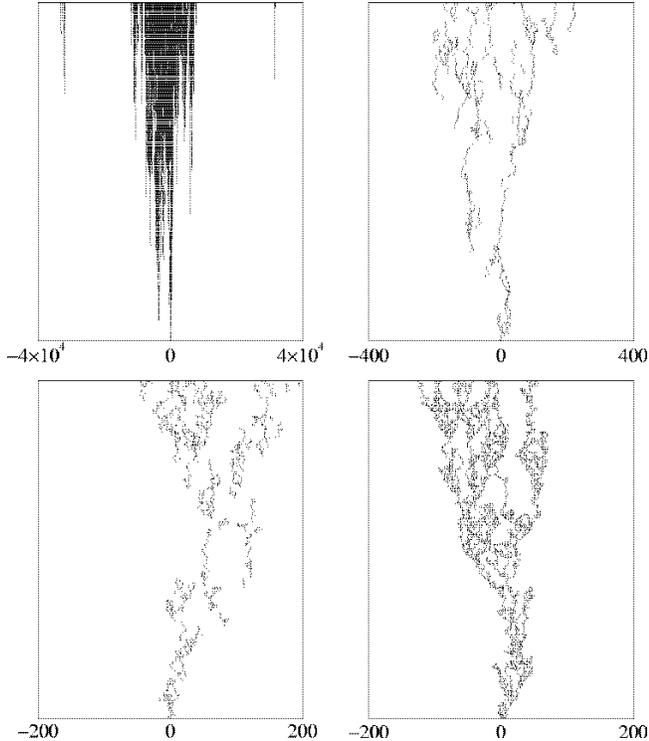}
}
\vskip 0.25 cm
\caption{Sample runs at various values of $\sigma$, with time
evolution running up the page. All runs are for $500$ time steps, and
are at values of $p$ about $10\%$ away from the critical point into
the active phase. The top two frames show $p=0.9$, at $\sigma=1.0$ and
$\sigma=1.5$, from left to right.  The lower left frame shows
$\sigma=1.9$, $p=0.77$, and the lower
right frame shows the ordinary short--ranged BARW model at
$p=0.46$. Notice the large change in scale between the first and last
frame.} 
\label{sample_runs}
\end{figure}

Two different initial conditions were used to calculate the 
different exponents. In the first case, the initial condition was a 
``seed'' of two particles at lattice sites $\pm 1$.  See
Fig.~\ref{sample_runs} for some sample runs with this initial
condition, run for $500$ time steps.  The long--range hops at small
$\sigma$ result in a very rapid and wide--ranging dispersal of the
particles. 

\begin{figure}
\centerline{\epsfxsize=8.5cm 
\epsfbox{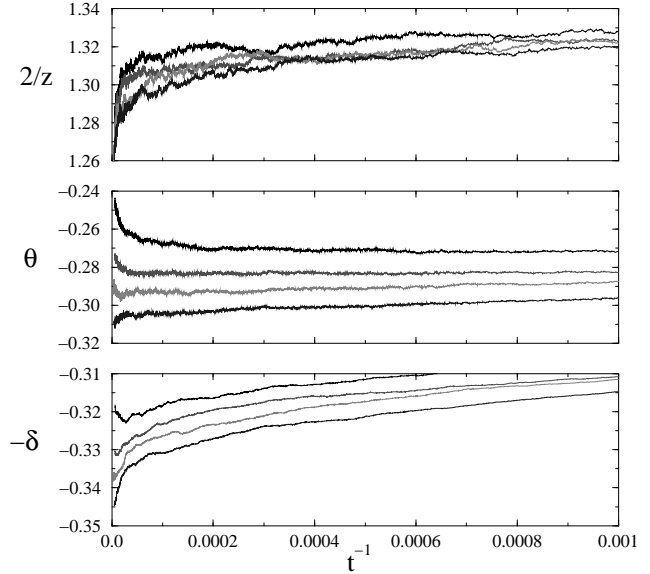}
}
\caption{The effective local exponents, as in
Eqs.~(\ref{local_delta})-(\ref{local_z}), for
$\sigma=1.6$.  The exponents are plotted
against $1/t$ to extract the $t\rightarrow \infty$ limit. The
curves correspond to values of $p$, from top to bottom, of $0.985$,
$0.988$, $0.99$, and $0.992$. }
\label{sample_exp}
\end{figure}

These simulations 
were averaged over many runs from the same initial condition but for
different sequences of random numbers. The number of runs, $P(t)$,
surviving to time $t$, the number of particles in the system averaged
over the total number of runs, $N(t)$, and a mean square spreading
distance, $R^2(t)$, were all measured. The
mean spreading distance is defined by a geometric mean in the L\'evy
case (see Ref.\cite{hinhow} for more details). At the critical point,
these quantities should all follow power law behavior with
\begin{eqnarray}
& & P(t) \sim t^{-\delta} , \label{local_delta} \\
& & N(t) \sim t^{\theta} , \label{local_theta} \\
& & R^2(t) \sim t^{2/z} . \label{local_z}
\end{eqnarray}
The critical point is determined by plotting a local exponent against
$1/t$, and estimating the value of $p$ which produces a straight line
as $1/t \rightarrow 0$. For the survival probability, the local
exponent is defined by  
\begin{equation}
-\delta(t) = {{\ln {{P(t)} \over {P({{t} \over {b}})}}} \over {\ln b}},
\end{equation}
and similarly for the other quantities. We have used $b=5$ in our data
analysis. The extrapolation of the exponent to $1/t\to 0$ is the
estimate of its long time value. A sample analysis, for $\sigma=1.6$,
is shown in Fig.~\ref{sample_exp}. 

We focus first on the perturbatively inaccessible transition found for
larger values of $\sigma>\sigma_c'(d=1)$. In this regime we performed
simulations at criticality and measured the exponents defined in
Eqs.~(\ref{local_delta})-(\ref{local_z}). To reduce finite size
effects, we implemented periodic boundary conditions and used a very
large lattice. Rather than store the occupation numbers of each
lattice site, only the positions of the particles were stored. This
meant that the system size was limited only by the number of integers,
i.e. a system size of $2^{64} \approx 1.8 \times 10^{19}$ on the
$64$--bit computer used. The simulations ran for times of between
$2 \times 10^4$ and $2 \times 10^5$ time steps per particle, and were
averaged over at least $2 \times 10^6$ runs. 

We encountered several obstacles in accurately determining the values
of these 
exponents. First, the quantities measured were also expected to behave
as power laws on the (critical) inactive side of the transition. The
difference between the equivalent exponents on the critical line and
in the critical phase were sometimes small, particularly near
$\sigma=\sigma_c'(d=1)$, where the fixed point we investigated merges
with the pure L\'evy annihilation fixed point. Consequently
measurements close to this point required the longest runs. Also,
corrections to scaling meant that the effective local exponents did,
in fact, vary with $t$. The impact of these corrections to scaling was
sometimes difficult to interpret accurately. Finally, the various
exponents measured yielded slightly different estimates for $p_c$,
making it difficult to determine the critical point more accurately
than done here. This may be due to the corrections to scaling
being of different sizes for each of the exponents measured.

\begin{figure}
\centerline{\epsfxsize=8cm 
\epsfbox{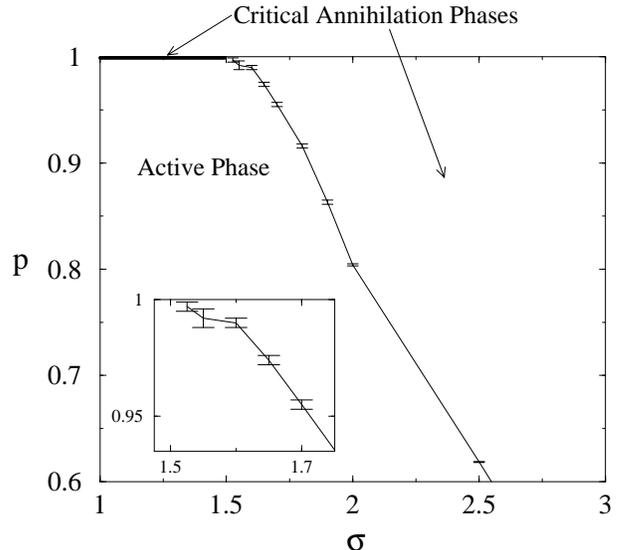}
}
\caption{Phase diagram for the BALF model in $d=1$.  The inset shows a
blowup of the region near $\sigma_c'(d=1)$. }
\label{phase}
\end{figure}

The results of these simulations are shown in Table~\ref{exp_table}.
The $\sigma = \infty$ values shown are for simulations with normal
diffusion, where the exponents measured are consistent with those of 
other
simulations~\cite{jensen:1994,kwon-park:1995}. Fig.~\ref{phase}
shows the phase diagram as determined by the simulations. 

We now discuss some features of the numerical data in
Table~\ref{exp_table}:

$\bullet$ The data presented in Table~\ref{exp_table} is consistent
with a value very close to $\sigma=\sigma_c'(d=1)=3/2$ for the
emergence of the critical L\'evy annihilation
phase at nonzero branching. This is in good agreement with the one
loop result for $y_{\mu}$ in Eq.~(\ref{ymu}), and provides some
evidence that this one loop result may in fact be exact, as it was for
the short--ranged L\'evy case. 

$\bullet$ The measured exponents changed by rather small amounts over
the range of $\sigma$ studied. As discussed in section~III, the
exponents at $\sigma=\sigma_c'(d=1)$, $p=1$, can be calculated for the
pure L\'evy annihilation model, and, assuming $\sigma_c'(d=1)=3/2$,
are given by $\delta=-\theta=1/3$, $z=3/2$. If the exponents are to
change monotonically as $\sigma$ is varied, then they are 
trapped in a relatively small range of values between the BARW and
L\'evy annihilation exponents.

$\bullet$ The numerical evidence also
strongly favors a smooth movement of the critical value $p_c$ away
from unity as $\sigma$ is increased above $3/2$, as shown in
Fig.~\ref{phase}. This finding has
consequences for the analogous short--ranged BARW model. In that case
the analog of the point at $\sigma=\sigma_c'(d=1)\approx 3/2$ is the
second critical
dimension found at $d_c'\approx 4/3$. The uncontrolled truncated loop
expansion used to analyze this point in Ref.~\cite{cardy-tauber}
predicted a discontinuous jump of the critical point as dimension $d$
is lowered through $d_c'$. The above numerical evidence argues against
this scenario and would rather predict a smooth movement of the
critical point. Given the uncontrolled nature of the truncated loop
expansion, any failure to accurately capture the behavior close to
$d_c'$ would not, perhaps, be very surprising. Nevertheless, our
results have, for the first time, provided numerical evidence for one
of the main conclusions of Ref.~\cite{cardy-tauber}, namely the
presence of a second critical dimension $d_c'$.

$\bullet$ Despite considerable effort, the data reported in
Table~\ref{exp_table} are unfortunately not precise enough to answer
the question: at what value of $\sigma$ do the L\'evy results
cross over to those of the short--ranged
BARW model? Regrettably, the situation from a theoretical
perspective is no clearer, due to the absence of any controlled field
theoretic methods in this regime. 

We now turn our attention to the second regime for $d=1$ BALF, 
that for $\sigma<\sigma_c'(d=1)\approx 3/2$. In this case, it is not
appropriate to perform simulations at criticality, since in that case
we would only be measuring the exponents of the pure L\'evy
annihilation model. Hence we have performed off--critical simulations
in an effort to measure $\beta_{\rm dens}$ as a function of
$\sigma$. In this case, we used a second initial condition, a
fully occupied lattice of size $L=10^7$. We then allowed the
number of particles to decay away until a steady--state was
reached. The steady--state density depends on the deviation from the
critical point, as described by Eq.~(\ref{n(Delta)}), and thus
$\beta_{\rm dens}$ may be directly measured. 

\begin{table}[t]
\begin{center}
\begin{tabular}{c|cccc}
$\sigma$ & $p_c$ &  $\delta$ & $\theta$ & $z$ \\ \hline
1.525& 0.997(2) & 0.32(1) & -0.30(1) & 1.53(2) \\
1.55 & 0.992(4) & 0.32(2) & -0.30(2) & 1.53(2)\\
1.6  & 0.990(2) & 0.33(2) & -0.30(2) & 1.56(2)\\
1.65 & 0.974(2) & 0.32(2) & -0.26(2) & 1.55(2) \\
1.7  & 0.955(2) & 0.32(2) & -0.24(2) & 1.59(2) \\
1.8  & 0.918(2) & 0.32(2) & -0.18(2) & 1.59(2)\\
1.9  & 0.863(2) & 0.32(2) & -0.14(1) & 1.63(2)\\
2.0  & 0.804(1) & 0.305(5) & -0.085(5) & 1.68(2) \\
2.5 & 0.6185(2) & 0.285(5) & -0.005(5)& 1.72(1)\\
$\infty$ & 0.5104(2) & 0.287(3)& 0.001(3) & 1.74(1)\\
\end{tabular}
\end{center}
\caption{The measured critical probabilities and exponents for various
values of $\sigma$. The number in brackets is an estimate of the error
in the last figure.}
\label{exp_table}
\end{table}

The values of $\beta_{\rm dens}$ measured in these steady--state
simulations are given in Table~\ref{beta_table}. For $\sigma<1$ the
mean field result should hold, since $d=1$ then lies above the upper
critical dimension $d_c=\sigma$. For $\sigma$ slightly bigger than
unity, the upper critical dimension will lie just above $d=1$, and
hence one might hope to directly observe the $\epsilon$ expansion
results (see also Ref.~\cite{hinhow} for a similar
case). Unfortunately, the values measured in the simulations deviate
by around $5-25\% $ from the mean field
or one loop $\epsilon$ expansion exponents calculated in
section~III. We believe there 
are two  reasons for this discrepancy. Firstly,
for small $\sigma$, finite size effects become important, as the
long--ranged hops allow a single particle to wrap all the way around
the system in a short time. Secondly, as $\sigma \rightarrow 
\sigma_c'(d=1)\approx 3/2$, $\beta_{\rm dens}$ becomes rather large,
and hence large systems and long runs were necessary to probe the very
small steady--state densities which occur near $p_c$.
Although we used as large a system as practicable, we were not able to 
entirely eliminate the discrepancy between theory and simulations. 

In summary, despite the difficulties encountered for
$\sigma<\sigma_c'(d=1)$, the overall picture that emerges from the
numerics agrees well with the theory presented
in the last section. As we have discussed earlier, this, in turn,
provides additional support for the analysis of the short--ranged BARW
model presented in Ref.~\cite{cardy-tauber}. 

\begin{table}
\begin{center}
\begin{tabular}{c|cc}
$\sigma$ & $\beta_{\rm dens}$ (measured) & $\beta_{\rm dens}$ (theory)
\\ \hline 
0.7 & 1.0 & 1 (mean field) \\
0.9 & 1.1 & 1 (mean field) \\
1.1 & 1.3 & 5/4 (one loop) \\
1.3 & 1.8 & 5/2 (one loop) \\
\end{tabular}
\end{center}
\caption{The exponent $\beta_{\rm dens}$ determined in simulations of
the steady--state density of a system of size $10^7$.}
\label{beta_table}
\end{table}

\section{Conclusions}

In this paper we have presented an analytic and numerical study of the
BALF model. Using field theoretic techniques we have obtained a good
analytic understanding of the model in the physical dimension $d=1$
for the regime $\sigma$ less than
about $3/2$. For values of $\sigma$ larger than this, we have had to
rely solely on numerical simulations. In both regimes the critical
exponents of the active/absorbing transition are found to vary
continuously with the L\'evy index $\sigma$. Numerically, we find
that the transition between the two regimes in $d=1$ occurs at
$\sigma=\sigma_c'(d=1)\approx
3/2$, in agreement with the one loop result from the field
theory. Unfortunately our numerics for the small $\sigma$ regime were
not good enough to confirm the accuracy of our $\epsilon$ expansion
calculations. Nevertheless this is the first time this universality
class has been accessed, since its equivalent in the original BARW
model lies in the inaccessible dimensions $d_c'<d<2$. 

Finally, we would like to emphasize that L\'evy flights are a powerful
way of probing the higher dimensional behavior of
nonequilibrium models while performing simulations only in $d=1$. The
disadvantage of this approach is that it necessitates the use of
extremely large system sizes if finite size effects are to be
avoided. However, we have shown that in large regions of parameter
space these problems can be overcome, and reasonable estimates
obtained for the exponents. 

\section*{acknowledgments}

We would like to thank John Cardy, Kent Lauritsen, and Uwe T\"auber
for very useful discussions. We also thank Mike Plischke for a
critical reading of the manuscript. We acknowledge support from 
NSERC of Canada.

\end{multicols}

\bigskip
\widetext
\begin{multicols}{2}

\end{multicols}

\end{document}